\documentclass[showpacs,floatfix,twocolumn,a4paper,prl]{revtex4}
\usepackage{amsmath}
\usepackage{epsf,graphicx}

\begin{document}

\title{Using isosbestic points to extract interactions from structure
factors } \author{A. A. Louis} \affiliation{Dept. of Chemistry,
University of Cambridge, Lensfield Road, CB2 1EW, Cambridge, UK}
\begin{abstract}
Inverting scattering experiments to obtain effective interparticle
interactions for particles in solution is generally a poorly
conditioned problem.  More accurate potentials can be obtained through
the use of {\em isosbestic points}, values of $k$ where the scattering
intensity $I(k)$ or the structure factor $S(k)$ is invariant to
changes in potential well-depth.  These points also suggest a new
extended corresponding states principle for particles in solution
based on the particle density or packing fraction, the second osmotic virial coefficient,
and a new measure of potential range.
\end{abstract}
%61.20.-p Structure of liquids
%82.70Dd Colloids
\maketitle

%\begin{multicols}{2}

Extracting effective interparticle interactions from experiments is a
central objective in chemical physics, because these interactions
govern physical behaviour\cite{Liko01}.  But like many such inverse
problems, this task is complicated. Experimental data is not perfect:
statistical fluctuations occur, and results may not be obtainable over
a complete parameter range.  Moreover, the inversion procedures are
often ill conditioned: small differences in the initial input can
result in large changes in the final output.  This letter concentrates
on a well-known class of such problems, namely the inference of
(spherically symmetric) effective interparticle pair potentials
$v^{eff}(r)$ from experimental scattering data for particles in
solution.  For that reason lower effective volume fractions $\eta$ are
investigated than in the better studied case of simple liquids near
the triple point, where the dominance of entropic hard-core
forces\cite{Hans86} makes it difficult to distinguish between
different attractive potentials\cite{Reat86,Hans86}.  Even though at
lower volume fractions attractive interactions can determine the
structure\cite{Loui01a}, this letter shows that inversions are equally
problematic, albeit for different reasons.

% The methodology typically resembles that of the better
%studied case of simple liquids near their triple point[ ], but for
%particles in solution the experimentally relevant packing fractions
%are often much lower.
% For the former systems the
%structure is primarily determined by the (entropic) hard-core
%correlations, while for the latter systems the short-range attractive
%(energetic) forces dominate the structure\cite{Loui01a}
%

For particles in solution, the scattering intensity $I(k)$, measured
by light, X-rays, or neutrons, is usually interpreted by dividing
$I(k)$ by the single particle form factor $P(k)$, to obtain the
structure factor $S(k) = I(k)/(\rho P(k))$\cite{Hunt01}; here
$\rho=N/V$ is the density. From $S(k)$, inferences are then made about the
form of the inter-particle interactions $v^{eff}(r)$. However, this
inversion is ill-conditioned: Within typical experimental accuracy, a
single $S(k)$ at a given state-point can be interpreted with a wide
range of different $v^{eff}(r)$'s.  But, as this letter
demonstrates, these problems can be circumvented by introducing a new
concept, {\em isosbestic points} -- values of the wave-vector $k$ for
which $S(k)$ is invariant under changes of the attractive potential
well depth.  Isosbestic points determine the effective range of the
$v^{eff}(r)$, and also suggest an extended corresponding states
principle for particles in solution, described by three experimentally
accessible variables: the
range defined by these points, the particle density $\rho$ and the
reduced second osmotic virial coefficient $B_2^*=B_2/B_2^{HS}$, where
$B_2^{HS}$ is the second virial coefficient of a hard-sphere (HS)
system.
\begin{figure}
\includegraphics[width=8cm]{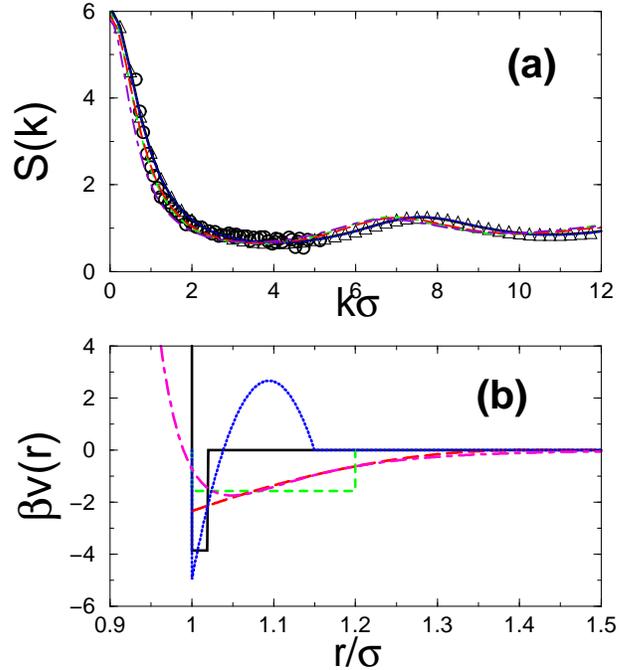}
\caption{\label{fig:Huang} A single set of experimental data can be
interpreted by many different effective potentials.
{\bf (a)} The experimental
SANS data (circles) on a microemulsion at packing fraction $\eta =
0.075$ is from \protect\cite{Huan84}.  The open triangles denote the
$S(k)$ from the Baxter model\protect\cite{Baxt68} for $B_2^* = -1.60$,
while the other $S(k)$ are calculated for the potentials
shown in {\bf (b)}. These include SWs with $\Delta_{SW}=0.02$ (solid
lines) and $\Delta_{SW}=0.2$ (dashed), an AO form with
$\Delta_{AO}=0.4$ (long-dashed), a generalised depletion form with
$\Delta_D=0.2$(dotted) and a LJ-12 (dot-dashed) potential.  The line-styles in
the two graphs correspond with each other. 
 }
\vspace*{-0.7cm}
\end{figure}

Fig.~\ref{fig:Huang} helps set the stage and introduce the problem to
be addressed here. Experimental data\cite{Huan84}, taken
from small-angle neutron-scattering (SANS) of a microemulsion composed
of small water droplets, coated with a layer of surfactant (AOT), is
compared to some theoretical structure factors $S(k)$.  In the
original letter\cite{Huan84} a best fit, using an approximate integral
equation technique, yielded a hard-core diameter of $\sigma = 60 \AA$,
a packing fraction $\eta = \pi \sigma^3/6 = 0.075$, and an attractive
square well (SW) potential with a range $\Delta_{SW} = 0.02 \sigma$ and a well
depth of $\beta v(\sigma) = 3.85$.  The quality of the integral
equation fit was then confirmed by independent computer simulations.
However, as strikingly demonstrated by Fig.~\ref{fig:Huang}, a wide
range of other potentials, depicted in Fig.~\ref{fig:Huang}(b), also
lead to fits of similar accuracy.  These include HSs with
 attractive SWs of two different ranges $\Delta_{SW}$,
an Asakura-Oosawa (AO) depletion potential\cite{Asak58} of range
$\Delta_{AO}$,  an alternate simplified depletion potential of
range $\Delta_D$ showing the effects of solvation layers\cite{Gotz98},
as well as a Lennard-Jones-n (LJ-n) potential, defined as:
\begin{equation}\label{eq2}
 v(r) = 4 \epsilon \left(\left(\frac{\sigma}{r}\right)^{2n} -
\left(\frac{\sigma}{r}\right)^{n}\right),
\end{equation}
with $n=12$.  The $S(k)$ are calculated within the Percus Yevick (PY)
integral equation closure\cite{Hans86,PY}, and are close to the
analytical PY solution of Baxter's model\cite{Baxt68}.

%Clearly, the inversion in Fig.~\ref{fig:Huang} is ill-conditioned.
%Since these were pioneering experiments, one might ask if it would be
%possible to significantly improve the accuracy and the range of the
%experimental data to be fit.  
Improving significantly on these pioneering experiments is hard.
$S(k)$ is obtained by dividing $I(k)$ through by $P(k)$, which is only
approximately determined, and which rapidly decays to zero for
increasing $k\sigma$, so that achieving better accuracies for a larger
$k\sigma$ range is very difficult. Moreover, there are few alternative
methods to directly measure the interparticle interactions.  In
contrast, for simpler atomic or molecular systems, gas-phase
experiments or ab-initio calculations can provide accurate independent
estimates of the interactions, and impose important constraints on the
potential forms with which to fit the experimental
data\cite{Maitland}.  Although increasingly accurate techniques being
developed for the real-space measurement of the $v^{eff}(r)$ of
particles in solution\cite{Isra92}, these are often limited to certain
particle types or sizes.

The upshot of Fig.~\ref{fig:Huang} is that deducing an effective
potential from experimental scattering data at one state point is
difficult; {\em the inversion is ill-conditioned}. Clearly more information
is needed to interpret the data\cite{critique}. 

But first, what  can be inferred from a single $S(k)$?  One hint comes the
Baxter model, which is completely determined by the packing fraction
$\eta$ and the reduced osmotic virial coefficient $B_2^*$ or equivalently, the
Baxter parameter, defined as $\tau = \frac14/(1- B_2^*)$.  The $S(k)$
in Fig.~\ref{fig:Huang}, with effective $\tau$ parameters ranging from $\tau =
0.093$ to $\tau = 0.10$   are all close to the Baxter model result with
$\tau = 0.096$.   
%Their effective $\tau$ parameters range 
% For most of these systems, the critical point is at a very
%similar $\tau$\cite{Vlie00,Noro00}, but at slightly higher $\eta$, so
%that the present $S(k)$ are expected to be near the spinodal.
  To first order therefore, measuring a single $S(k)$ in
this regime of low packing fraction, commonly encountered for
particles in solution, does not result in much more information about
the effective potential than the reduced second virial coefficient
$B_2^*$. Just as in simple liquids inversions are hard because the
$S(k)$ resemble those of HSs\cite{Hans86}, here difficulties arise because they
are close to  the Baxter $S(k)$.

\begin{figure}
\includegraphics[width=8cm]{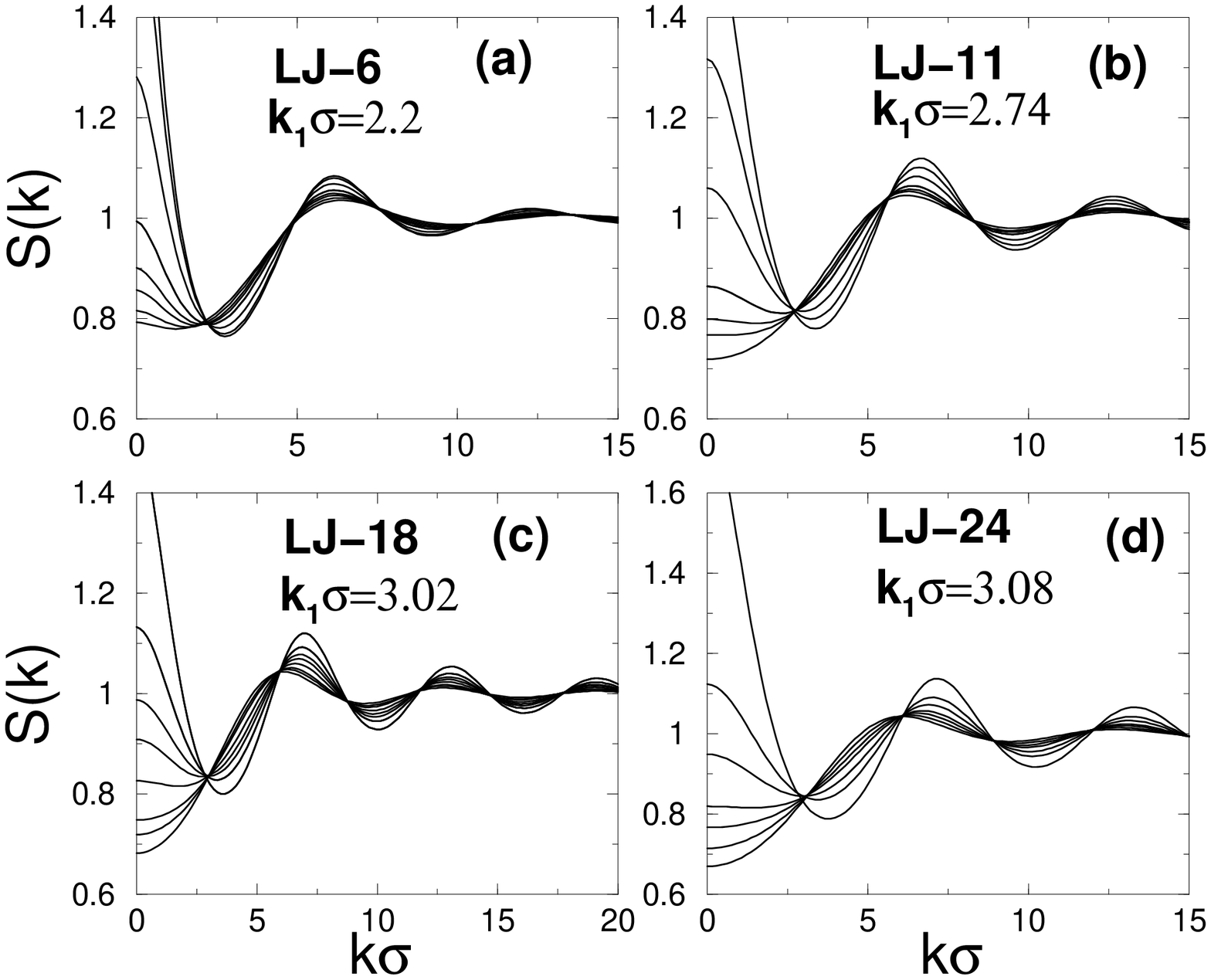}
\includegraphics[width=8cm]{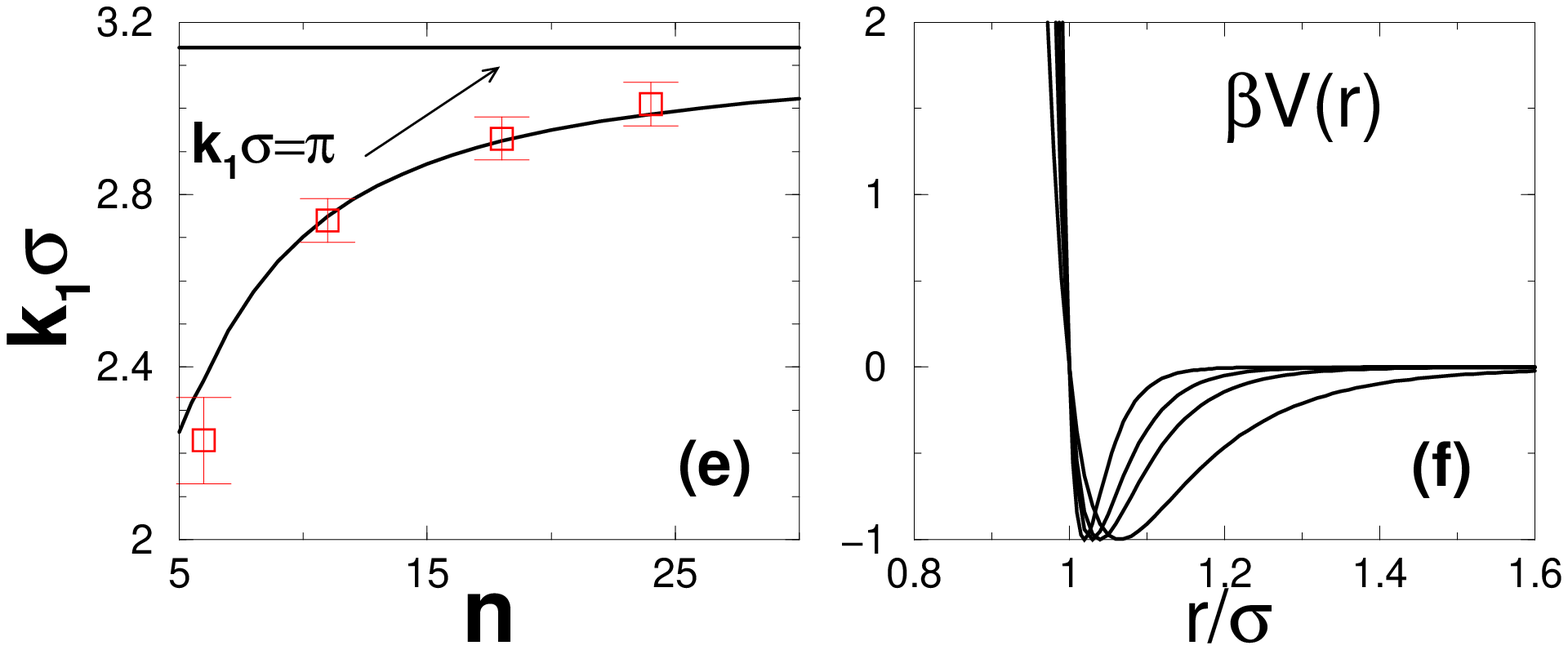}
\caption{\label{fig:Isosbestic}(a)-(d) $S(k)$ were calculated at $\eta
= 0.0576$ and at different temperatures for each of the four LJ-n
potentials $v^{eff}(r)$ given by Eq.~(\protect\ref{eq2}).  Different
$v^{eff}(r)$ lead to different isosbestic points $k_n\sigma$.  (e) The
first isosbestic point $k_1\sigma$ approaches $\pi$ with increasing
$n$. A simple theory (solid line), described in the text, fits the
data well. (f) The range of the  LJ-n potentials  in (a)-(d) 
decreases with increasing $n$.\vspace*{-0.4cm}}
\end{figure}

To make further progress, measurements at other state points are
necessary.  This is done in Figs.~\ref{fig:Isosbestic}(a)-(d), which
depict the $S(k)$ calculated for four LJ-n potentials at different
temperatures.  A best fit to the Baxter model $S(k)$ at each
temperature would result in $B_2^*$ as a function of temperature.  But
there is clearly more information in these curves: Each of the four
$v^{eff}(r)$ results in a different set of {\em isosbestic points},
where the $S(k)$ is invariant for different temperatures.
Fig.~\ref{fig:Isosbestic}(e) shows that the first isosbestic point
$k_1\sigma$ increases with increasing $n$, reflecting the decrease of
the range of the LJ-n potential~(\ref{eq2}) depicted in
Fig.~\ref{fig:Isosbestic}(f). (Error bars in
Fig~\ref{fig:Isosbestic}(e) reflect the approximate nature of 
the isosbestic points.)

These observations can be quite easily rationalised by the following
simple theory: If the total correlation function $h(r)=g(r)-1$ is
split into two parts, with $h_0(r) = -1$ for $r < \sigma$, and $h_1(r)
= g(r)-1$ for $r \geq \sigma$, then the structure factor $S(k) = 1 + \rho \hat{h}(k)$ simplifies to
\begin{equation}\label{eq3}
S(k) = 1 + \frac {4 \pi \rho}{k} j_1(k) + \rho \hat{h}_1(k),
\end{equation}
where $j_1(k)$ is the first spherical Bessel function, and
$\hat{h}_1(k)$ is the Fourier Transform (FT) of $h_1(r)$.  For
potentials with a hard-core, such as most of those depicted in
Fig.~\ref{fig:Huang}(b), this is exact, but even for the LJ-n
potentials this is a good approximation\cite{sigma}.  Since
$\hat{h}_0(k) = \frac{4 \pi \rho}{k} j_1(k)$ is
independent of temperature (barring small effective $\sigma$ effects),
Eq.~\ref{eq3} suggests that isosbestic points occur whenever
$\hat{h}_1(k)=0$.

\begin{figure}
\includegraphics[width=8cm]{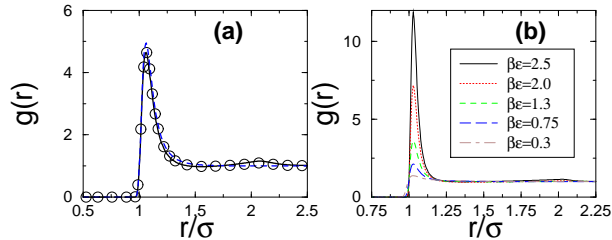}
\caption{\label{fig:gofr} (a) The PY approximation (solid line) very
accurately reproduces these molecular dynamics simulations\cite{PY} of
a LJ-12 potential at $\beta \epsilon = 1.6$, and $\eta=0.1$.  The
simpler $g(r) = \exp (-\beta v(r))$ form (dashed lines) is also
accurate. (b) When the temperature is changed, the dominant effect on
$g(r)$, calculated here with PY, is to change its amplitude.  }
\vspace*{-0.5cm}\end{figure}

It has already been pointed out that in many experimentally relevant
cases with low $\eta$, the $g(r)$ are surprisingly well approximated
by a simple form $g(r) = \exp (- \beta v(r))$\cite{Loui00a,Loui01a},
as explicitly demonstrated in Fig.~\ref{fig:gofr}. This
simple approximation, equivalent to taking a Mayer
function\cite{Hans86} for $h(r)$, works best for particles with short
range attractive potentials. These can become quite deep, leading to
large values of $g(r)$ near contact, well before the system crosses a
liquid-liquid or liquid-solid phase-line.  Within this approximation,
the dominant effect of varying the temperature is to change the
amplitude of $h_1(r)$ (as demonstrated in Fig.~\ref{fig:gofr}(b)). To
first order, the period of $\hat{h}_1(k)$ is not affected and each
$k\sigma$ where $\hat{h}_1(k)=0$ leads to an isosbestic point. For an
infinitely narrow potential, the isosbestic points would be at $k_n
\sigma = n\pi$, but for a finite potential range there is a
phase-shift, which explains why the $k_n\sigma$ move progressively
further from $n \pi$ with increasing range.
% The same holds for the
%higher order isosbestic points $k_n \sigma $, which are each shifted
%from a value $n \pi$.
  The excellent accuracy of this simple Mayer function theory for the first
isosbestic point $k_1\sigma$ is
demonstrated by the solid line in Fig.~\ref{fig:Isosbestic}(e).
% The largest deviation is for the
%longest range potential, which is not surprising, since the critical
%point is at a relatively lower temperature, which means that the
%maximum amplitude of $g(r)$ is less than for shorter range potentials
%so that the simple assumption that changes in temperature only change
%the amplitude but not the shape of $g(r)$ is least accurate here.

\begin{figure}
\includegraphics[width=8cm]{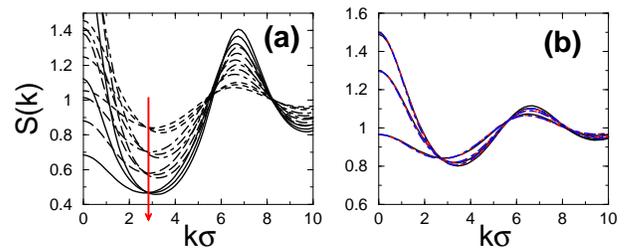}
\caption{\label{fig:sofk}Fig (a) The position of isosbestic points
barely varies with density for LJ-12, shown here for three
temperatures $\beta \epsilon = 1.5,1.33,1.09$ at each density
$\eta=0.05,0.1,0.15,0.2$. The line with the arrow denotes $k_1\sigma$
for increasing $\eta$.  (b) Three potentials leading to the same isosbestic
points: SW ($\Delta_{SW} = 0.26$) (solid lines), AO ($\Delta_{AO}=0.6$)
(dashed lines) and LJ-22 (dot-dashed lines) for $\eta=0.05$ and
$B_2^*=0, -0.719, -1$ (plots in descending  order of $B_2$ at $k=0$). }
\vspace*{-0.5cm}\end{figure}

The examples depicted in Fig.~\ref{fig:Isosbestic} are at a relatively
low packing fraction, but the isosbestic points are robust up to
packing fractions of at least $\eta = 0.2$, as shown in
Fig.~\ref{fig:sofk}(a)\cite{density}. These points are accurately
described by the simple theory, described above.
% which is equivalent
%to the FT of the Mayer function.

%The isosbestic points in Fig.~\ref{fig:Isosbestic} were at a
%relatively low $\eta$, which is also the case for many experiments.
%As the $\eta$ increases, one might expect the simple theory above to
%begin to break down, but remarkably, isosbestic points remain stable
%up to packing fractions of at least $\eta = 0.2$, as shown in
%Fig.~\ref{fig:sofk}(a)\cite{density}.

%The structure factors in Figs.~\ref{fig:Isosbestic} and \ref{fig:sofk}
%suggest that one could use the isosbestic points to determine 
%the range of  effective potentials from scattering experiments.
%However, the question still remains: What is meant by the range?  For
%individual potentials (like the SW or AO) it may be clearly defined,
%but how does one compare it between different forms?

%Here again, isosbestic points provide an interesting clue.
%Fig.~\ref{fig:sofk}(b) shows that at low densities the $S(k)$ for
%three different potentials, selected to have the same isosbestic
%points, show virtually identical structure for each value of $B_2^*$.
%In other words, given the isosbestic point $k_1 \sigma$, and the
%reduced virial coefficient $B_2^*$, the $S(k)$ are almost completely
%determined.

%As mentioned before, the $S(k)$ of simple atomic liquids near the
%triple point, are often very close to those of HSs.  But for the
%$S(k)$ discussed in this letter, that is manifestly not the case. For
%example, $\lim k \rightarrow 0 S(k) \leq 1$ for HS.  Instead it
%appears that the structure factors are much closer to a universality
%class based on the Baxter model (see e.g.\ Fig.~\ref{fig:Huang}).

If varying a parameter leads to isosbestic points, then this suggests
that the well-depth of $v^{eff}(r)$ is changing while the range is
not. When combined with the variation of $B_2$, this can help fix the
form of $v^{eff}(r)$.  However, this information may still not be
sufficient, since the well-depth of $v^{eff}(r)$ may depend in an
(unknown) non-linear fashion on the parameters like the temperature
$T$, the pH, and salt or other additive concentration.  Furthermore, as
shown in Fig.~\ref{fig:sofk}(b), different potential shapes, picked to
have the same isosbestic points, can generate similar $S(k)$ for each
value of $B_2^*$.

\begin{figure}
\includegraphics[width=8cm]{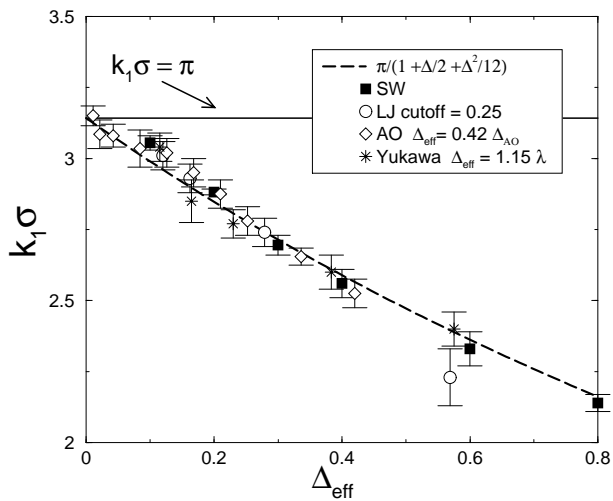}
\caption{\label{fig:univ} Defining effective range parameters from
isosbestic points. For each potential type a simple linear mapping to
an effective square well range $\Delta_{eff}$ was used (as described
in the text). The long-dashed line denotes an accurate analytical
approximation for the isosbestic points of square-well fluids which
can be used to estimate the $\Delta_{eff}$ for different potentials
directly from the $k_1 \sigma$.  }
\vspace*{-0.5cm}\end{figure}

This similarity of the $S(k)$ ties in with the work of Noro and
Frenkel (NF), who proposed that many properties of particles in
solution could be understood from an an extended corresponding state
principle based on the variables $\eta$, $B_2^*$, and an effective well
depth $\beta \epsilon$\cite{Noro00}.  However, the three potentials
shown in Fig.~\ref{fig:sofk}(b) do not have the same well-depth, while
still showing similar $S(k)$.  This suggests an alternative
corresponding states principle based on the variables $\eta$, $B_2^*$,
and an effective range $\Delta_{eff}$ related to the isosbestic
points. In fact, as NF point out, the potential range is not always
uniquely defined when comparing different $v^{eff}(r)$'s. They
suggested a non-linear mapping based on $B_2^*$ to derive an effective
SW range for each potential they consider.  I used a simpler linear
mapping to define the range $\Delta_{eff}$ of a given potential as
identical to that of a SW with the same $k_1 \sigma$.  For the LJ-n
potentials this criterion reduces to the distance $r-\sigma$ where
$\beta v(r)$ reaches $1/4$ of its maximum depth, for the AO potential
the effective SW range is $\Delta_{eff} \simeq 0.42 \Delta_{AO}$, and
for the hard-core Yukawa potential, with an attractive tail of the
form $ v(r) = \epsilon \exp(-r/\lambda)$ for $r>\sigma$, the mapping
is $\Delta_{eff} \simeq 1.15 \lambda$.  The results are depicted in
Fig.~\ref{fig:univ}, and show that these simple linear mappings
successfully define an effective range for each potential type.  An
accurate approximation of $k_1 \sigma$ for the (effective) SW is given
by $k_1 \sigma \approx \pi/(1 + \Delta_{eff}/2 + \Delta_{eff}^2/12)$,
which follows from an expansion of the exact solution to the simple
theory of isosbestic points.

Another argument for including the $\Delta_{eff}$ (instead of the
well-depth) as one of the relevant $3$ variables is that the range
determines the topology of the phase-diagram\cite{Noro00}.  The
critical range below which the fluid-fluid transition becomes
metastable to the fluid-solid line is at $\Delta_{eff} \approx 0.15$
for all four potentials (SW, AO, Yukawa and LJ-n), something
consistent with what NF found.  In addition, $\Delta_{eff}$ is
directly experimentally accessible through $k_1\sigma$.  For example,
taking advantage of the proximity of a metastable fluid-fluid critical
point to enhance nucleation, a proposal recently made for protein
solutions\cite{tenW97}, would involve tailoring solution conditions
such that the measured isosbestic point is just above $k_1\sigma =
2.92$, so that $\Delta_{eff}$ is just below $0.15$.

% A more detailed comparison of the two
%schemes for defining an extended corresponding state description will
%be discussed in a later publication.

In summary then, extracting effective potentials $v^{eff}(r)$ from
$S(k)$ at the lower packing fractions typically encountered for
particles in solution is a poorly conditioned problem. Whereas in
simple liquids the HS model describes the dominant features of the
$S(k)$, here the Baxter model appears to be the fundamental underlying
model for $S(k)$ around which different attractive  potentials  only induce
a mild perturbation. More information can be obtained by studying the
isosbestic points, which  help determine the range $\Delta_{eff}$ of the
potentials. Together with $\eta$ and $B_2^*$, $\Delta_{eff}$ can be
used to define an extended corresponding states principle.  Particles
in solution with similar values of these three parameters should have
similar properties, such as the relative stability of the fluid-fluid
and fluid-solid binodals.  In other words, many solution properties
could be deduced without having to actually invert to an explicit form
of $v^{eff}(r)$.

Another advantage of a description based on these variables is that
they are  directly experimentally accessible.  In fact, numerous examples of
isosbestic points can be found in the literature, ranging from
colloids with short-range sticky coats\cite{Duit91}, to
colloid-polymer mixtures\cite{Ye96}, to globular
proteins\cite{Tard99}, to magnetic colloids\cite{Dubo99} and to
colloid-micelle mixtures\cite{Bart}. Interestingly, these points
should also appear in $I(k)$, although determining  $\Delta_{eff}$ still depends on an accurate $P(k)$.
Future work will consider the role of size polydispersity (expected to
mainly affect higher order $k_n\sigma$) and the existence of
isosbestic points for non-spherical objects.

The author thanks P. Bartlett for first bringing isosbestic points to
his attention, J. P. K. Doye for helpful discussions, and The Royal Society for financial support.


\begin{thebibliography}{99}
\bibitem{critique} I should add that this is not an implicit critique of
ref\cite{Huan84}, where other physical considerations were indeed used
to infer the suggested short ranged potential.
\bibitem{sigma} There are several similar ways to define an effective
$\sigma_{eff}$ for a LJ-n potential, see e.g.\protect\cite{Hans86}.
A more careful analysis shows that for our examples, these don't differ much from
$\sigma$, and  have a marginal effect on the isosbestic points.
\end{thebibliography}
\end{document}